\documentclass[pra,showpacs,floatfix,amsmath,amsfonts,twocolumn]{revtex4}

\usepackage{graphics}
\usepackage{epsfig}
\usepackage{color}
\usepackage{eurosym}
\usepackage[normalem]{ulem}
\usepackage[makeroom]{cancel}

\newcommand{\PT}{{\cal PT}}

\newcommand{\chitwo}{\chi^{(2)}}

\begin{document}

\title{Solitons in a $\PT$- symmetric $\chi^{(2)}$ coupler}

\author{Magnus \"{O}gren$^1$}
\author{Fatkhulla Kh. Abdullaev$^2$}
\author{Vladimir V. Konotop$^3$}





\address{$^1$School of Science and Technology, \"{O}rebro University, 701 82 \"{O}rebro, Sweden\\ $^2$Physical-Technical Institute, Uzbekistan Academy of Sciences, 100084, Buyuk Ipak Yuli str, 2-b, Tashkent, Uzbekistan\\ $^3$Centro de F\'isica Te\'orica e Computacional and Departamento de F\'isica, Faculdade de Ci\^encias,
Universidade de Lisboa, Campo Grande 2, Edif\'icio C8, Lisboa 1749-016, Portugal}


\begin{abstract}
We consider the existence and stability of solitons in a $\chitwo$ coupler. Both the fundamental and second harmonics undergo gain in one of the coupler cores and are absorbed in the other one. The gain and losses are balanced creating a parity-time ($\PT$) symmetric configuration. We present two types of families of $\PT$-symmetric solitons, having equal and different profiles of the fundamental and second harmonics. It is shown that gain and losses can stabilize solitons. Interaction of stable solitons is shown. In the cascading limit the model is reduced to the $\PT$-symmetric coupler with effective Kerr-type nonlinearity and balanced nonlinear gain and losses.
\end{abstract}



\maketitle

Optical solitons in media with quadratic ($\chi^{(2)}$) nonlinearities was subject of intensive investigations over the last few decades both theoretically and experimentally~\cite{review1,review2}. 
Several types of quadratic bright soliton solutions have been reported in exact analytical form~\cite{KaramzinSukhorukov,WD94a,WD94b,MST94} and families of the solutions were investigated numerically~\cite{BK95}.   
Spatial one-dimensional quadratic solitons in optical waveguides have been observed~\cite{SBS96,SBSS99}. 

Direct nontrivial generalization of a guiding structures for the $\chitwo$ solitons~\cite{review1} towards their manipulations is a $\chitwo$ coupler. 
 Such device supports propagation of four different field components, which are two fundamental fields (FFs) and the respective second harmonics (SHs), in each of the two coupler arms. 
The coupling of the fields in the arms can be implemented in different ways. 
The simplest model was with the tunnel coupling between FF only, used for investigation of discrete $\chitwo$ solitons~\cite{Iwanow}. 
While carrier wave states and all-optical switching in  $\chitwo$ couplers was subject of many studies~\cite{Stegeman}, solitons in coupled optical waveguides with quadratic nonlinearities were explored much less. Numerical simulation of  propagation of temporal solitons and their switching  in the $\chitwo$ coupler has been performed in~\cite{Polyak}, while the existence of solitons and their stability for the case of no walk-off and full matching was shown in~\cite{mak97}.

\begin{figure}[ht!]
	\centering
		\includegraphics[width=0.55\columnwidth]{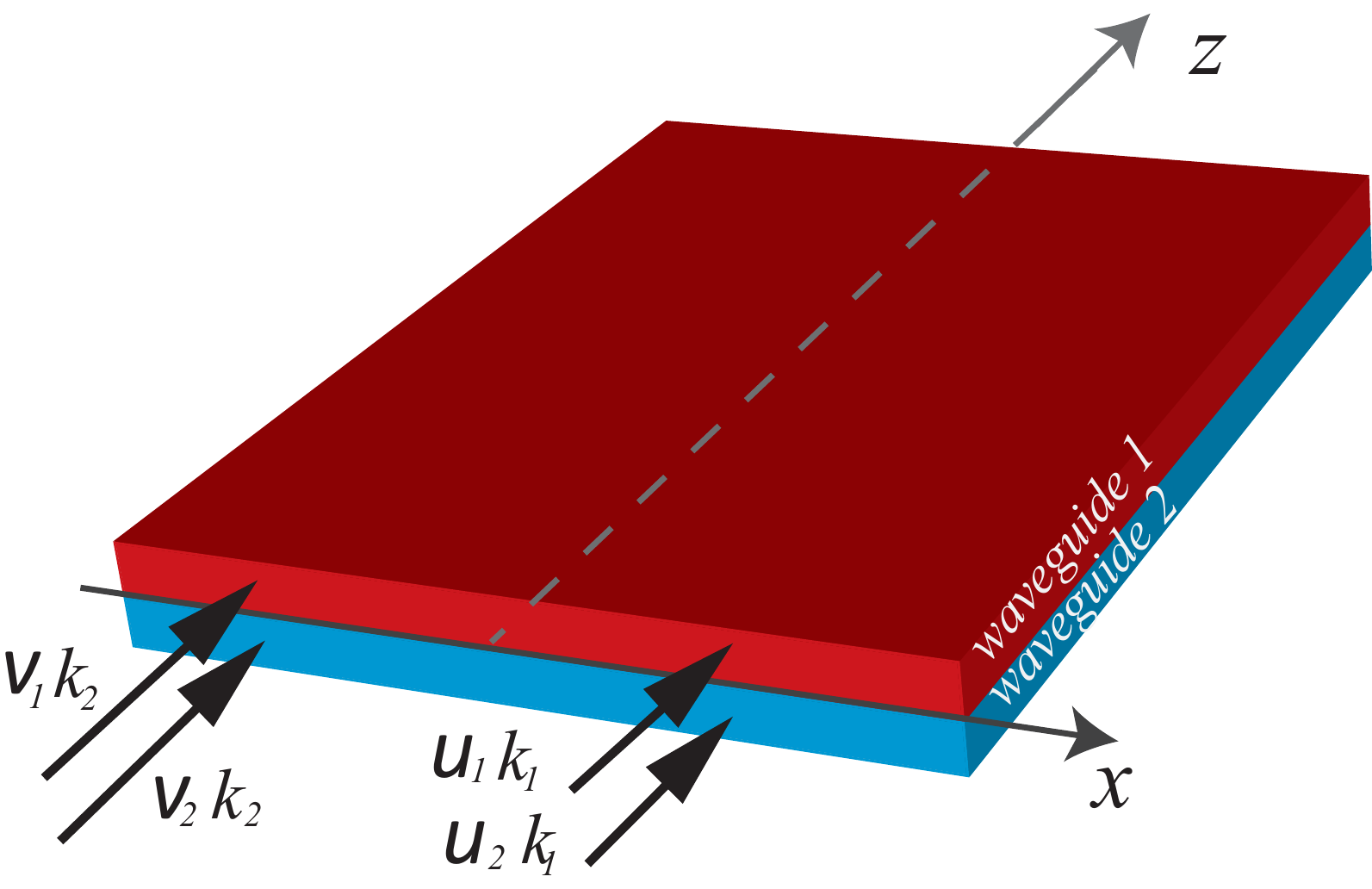}
	\caption{Schematic presentation of a planar $\chi^{(2)}$ coupler. The first and second waveguides have gain and absorption respectively. Four beams are applied at the input ($z=0$).
	}
	\label{fig:zero}
\end{figure}

In this Letter we investigate solitons in a $\chitwo$ coupler with gain in one arm and absorption in another one (as illustrated in Fig.~\ref{fig:zero}). 
The gain and loss are balanced, thus implementing a parity-time ($\PT$) symmetric~\cite{Bender} system. 
Motivation of our study resides in peculiarities of such a device. 
Indeed, in spite of gain and losses it allows for propagation of soliton families~\cite{review}, which depend on one (or several) parameters. 
Since the gain, usually implemented in a form of active impurities is controlled by an external pump field, the parameters of solitons can be varied at fixed parameters of the hardware, making the control flexible, and opening possibilities, for instance for novel types of optical switching or nonreciprocal devices. Furthermore, including gain and loss in the system changes the parameter regions of the existence and stability of $\chitwo$ solitons. 
Additionally, the cascading limit of such a coupler gives origin to a $\PT$-symmetric coupled nonlinear Schr\"odinger equations, of a new type.
Recently, exploring different settings it was found~\cite{Tripen,El,Sukho} that  gain and losses modify the matching conditions, making it possible the resonant mode interaction which otherwise is not allowed in the conservative waveguides.

Since four different harmonics are involved, from the theoretical point of view the $\chitwo$ coupler can be viewed as a particular type of the nonlinear $\PT$-symmetric "quadrimer". 
For the Kerr-type nonlinearity
quadrimers received considerable attention (see e.g.~\cite{review} and references therein).  
In the case  of $\chitwo$ nonlinearity the previous studies were restricted to stationary (nondiffractive) propagation~\cite{chi2sol}.

We focus on  interaction of waves occurring in coupled active and absorbing planar waveguides~(Fig.~\ref{fig:zero}). 
The equations describing light propagation in such $\chitwo$ coupler read:
\begin{eqnarray}
\label{basic_equations}
\begin{array}{l}
i{u}_{1,z} =  -  u_{1,xx}+ \kappa_1 u_2-2u_1^* v_1 + i\gamma_1 u_1,\\
i{v}_{1,z} =  - \frac 12 v_{1,xx}+ \kappa_2 v_2-u_1^2-q v_1 + i\gamma_2 v_1,\\
i{u}_{2,z} =  - u_{2,xx}+ \kappa_1 u_1-2u_2^* v_2 - i\gamma_1 u_2,\\
i{v}_{2,z} = - \frac 12 v_{2,xx} +\kappa_2 v_1-u_2^2-q v_2 - i\gamma_2 v_2.
\end{array}
\end{eqnarray}
Here we use the dimensionless variables~\cite{review2}
$
u_{j}=2L_d d E_{j}$ and $v_{j} = L_d d e^{iq z}E_{j}$, where $j=1,2$ labels the waveguide, for the electric field envelopes $E_j$ of the FFs ($u_j$) and SHs ($v_j$), the dimensionless propagation  distance $z= Z/2L_d$ and the transverse coordinate $x=X/\eta$, the linear coupling between harmonics in different arms  $\kappa_i=K_i L_d$ and the mismatch of the propagation constants   $ q=\Delta k L_d,$ assumed the same in both arms, where $K_{1,2}$ are the physical couplings of FF and SH (we consider them positive), $d=\omega_1/(\epsilon_0n_1 c)\chi^{(2)}$ is the parameter of quadratic nonlinearity, $\Delta k=2k_1 - k_2$ is the phase mismatch, $\eta$ is the characteristic beam width, and $L_d = k_1\eta^2$ is the linear diffraction length.
The strength of the gain in the first waveguide and absorption in the second waveguide are characterized by the parameters $\gamma_{j}>0$, for the first ($j=1$) and second ($j=2$) harmonics, respectively. 
The equality of  the gain and losses ensures $\PT$-symmetry of the coupler. 

The model (\ref{basic_equations}) includes diffraction effects and generalizes the model of the $\PT$-symmetric coupler considered in~\cite{chi2sol}. 
On the other hand inclusion of gain and losses represents a $\PT$-symmetric generalization of the conservative coupler considered in~\cite{mak97}. 
We also notice that the system~(\ref{basic_equations}) obeys Galilean invariance. 
Thus having found localized beams, which we described by the four-component vector  $\psi(x,z)=(u_{1}(x,z),v_{1}(x,z),u_{2}(x,z),v_{2}(x,z))^T$ with $T$ standing for the transpose, and which propagate along the $z$ direction, one readily obtains beams propagating under a nonzero angle $\theta$ with respect to the $z$-axis in the form 
$\left(e^{i\varphi}u_{1}(\xi,z),e^{2i\varphi}v_{1}(\xi,z),e^{i\varphi}u_{2}(\xi,z),e^{2i\varphi}v_{2}(\xi,z)\right)^T,$ where $\varphi= wx/2-w^2z/4$, $\xi=x-wz$ and $w=\tanh \theta$.
In the presence of gain, a necessary condition for the possibility of observing localized nonlinear beams is the stability of the zero solution. This corresponds to the choice of the parameters in the so-called  unbroken $\PT$-symmetric phase~\cite{Bender}. 
Such stability is obtained from the linear dispersion relation. Using the ansatz $u,v\sim e^{ibz+ikx}$  in (\ref{basic_equations}) we obtain four branches of the linear modes:
\begin{eqnarray}
 b_{1,2}=-k^2\pm\sqrt{\kappa_1^2-\gamma_1^2},
 \quad
 b_{3,4}=-\frac 12 k^2+q\pm\sqrt{\kappa_2^2-\gamma_2^2}.
\end{eqnarray}
Thus the $\PT$-symmetry is unbroken (real $b_j$:s)  if $\gamma_1<\kappa_1$ and  $\gamma_2<\kappa_2$. 
Below we restrict the discussions to these constraints.

Let   the  parameters satisfy  the relation as follows
\begin{eqnarray}
\label{special_case}
\kappa_1^2\gamma_2=2\kappa_2\gamma_1\sqrt{\kappa_1^2-\gamma_1^2}.
\end{eqnarray}
In this case one can introduce $\delta$ through the relations $\sin (\delta)=\gamma_1/\kappa_1,$ where $0 \leq  \delta \leq \pi/2$,
and define $\beta_1=\kappa_1\cos(\delta)$, and  $\beta_2=\kappa_2\cos(2\delta)$.
Then the system (\ref{basic_equations}) has a solution of the form (by analogy with the ansatz introduced in~\cite{DrbMal}):
$u_1= u,$  $u_2=\pm e^{\pm i\delta}u$,
$
v_1= v$, $v_2=\pm e^{\pm 2i\delta }v$,
where the functions $u$ and $v$ solve the standard system of the $\chi^{(2)}$ equations:
\begin{equation}
\label{reduced_equations}
i{u}_{z} = -   u_{xx}+\beta_1u -2u^* v,\,\,\,
i{v}_{z} = - \frac{1}{2} v_{xx}+ (\beta_2 -q)v-u^2.
\end{equation}
 
Our main goal is the analysis of the effect of the interplay between nonconservative terms and coupling between the two systems of solitons governed by~(\ref{basic_equations}). 
For the analysis of stationary solutions, we first define the total energy flow in the $j$-th waveguide $P_j=\int\left(|u_j|^2+2|v_j|^2\right)dx$. 
In the conservative case ($\gamma_1=\gamma_2=0$) the total energy $P=P_1+P_2$ is constant along propagation. 
In the presence of gain and losses it is generally not so any more and one computes
\begin{eqnarray}
\label{P}
\frac{dP}{dz}=2\gamma_1\int\left(|u_1|^2-|u_2|^2 \right)dx+4\gamma_2 \int\left(|v_1|^2-|v_2|^2 \right)dx.
\end{eqnarray}
This relation means that for a {\em stationary} solution, i.e. the solution of the form $u_j=\tilde{u}_j(x)e^{i\beta z}$ and $v_j=\tilde{v}_j(x)e^{2i\beta z}$,  the difference in the energy flows is defined by
\begin{eqnarray}
P_1-P_2=\left(1-\gamma_1 / \gamma_2\right)\int\left(|u_1|^2-|u_2|^2\right)dx.
\end{eqnarray}
Thus,  if $\gamma_1\neq\gamma_2$ (what corresponds to the most typical situation) the equality of the energy flows in two different waveguides requires $|u_1|=|u_2|$ and $|v_1|=|v_2|$.
Further, we notice that due to the $\PT$ symmetry, if a column-vector 
$\psi=(u_1(x,z),v_1(x,z),u_2(x,z),v_2(x,z))^T$  is a solution of (\ref{basic_equations}), then the $\PT$-transformed field $\tilde{\psi}=\PT\psi=(u_2^*(x,-z),v_2^*(x,-z),u_1^*(x,-z),v_1^*(x,-z))^T$
is also a solution. 
Thus a stationary $\PT$-symmetric solution, which is defined by the relation $\PT\psi=\psi$ (or more generally $\PT\psi=e^{i\vartheta}\psi$, where $\vartheta$ is a constant phase) supports the equality of the power flows. On the other hand if for a given solution $P_1\neq P_2$, then the obtained solution is non-$\PT$-symmetric.

\begin{figure}[ht!]
	\centering
	\includegraphics[width=0.8\columnwidth]{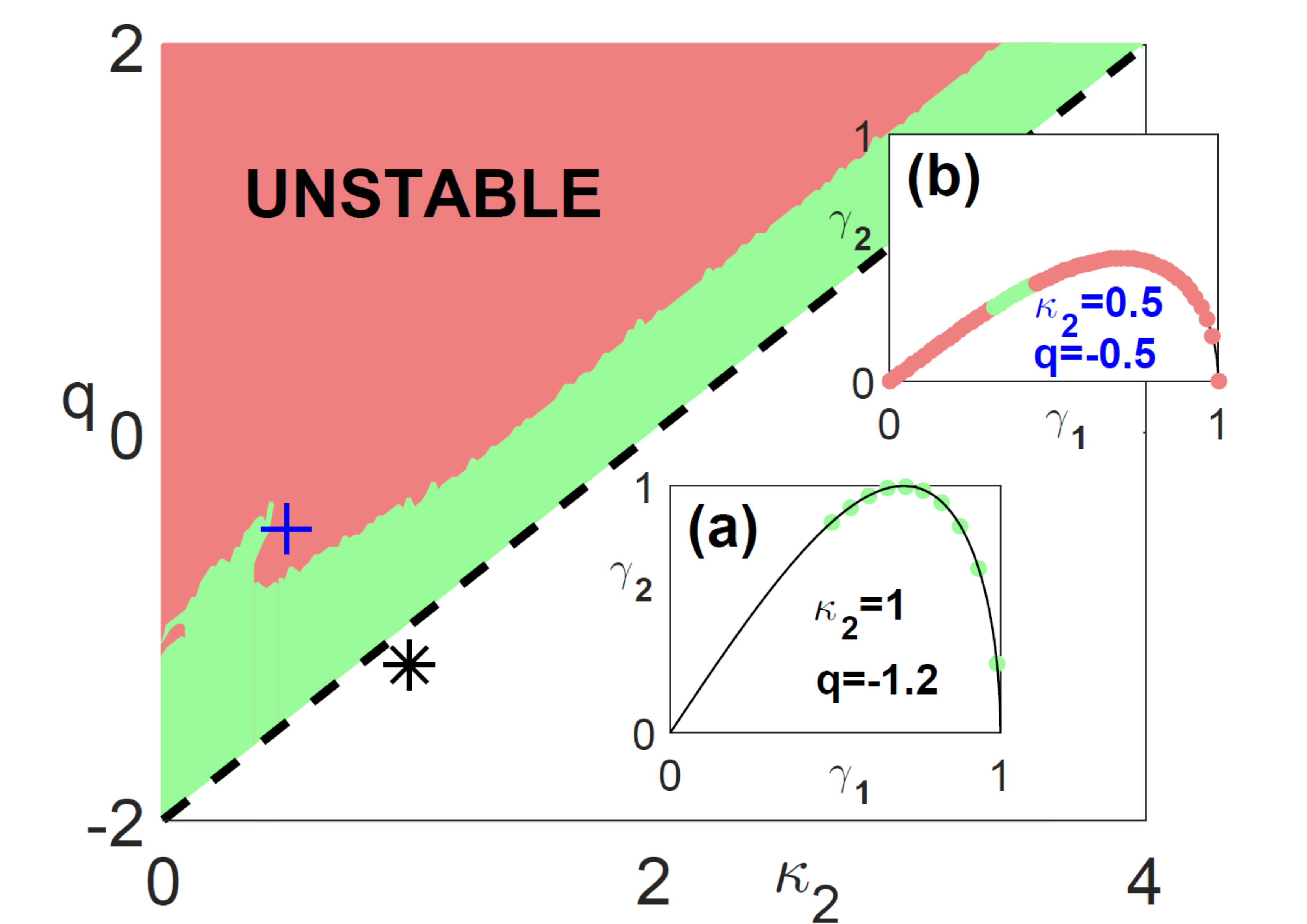}
	\caption{Stability for solitons~(\ref{KarSukh}) with equal intensity profiles of the FF and SH. 
		The main panel shows the conservative case ($\gamma_1=\gamma_2=0$).
		The inset plots are for specific values of $\kappa_2$ and $q$, where $0 \leq \gamma_1 < \kappa_1 = 1$ and $0 \leq \gamma_2 < \kappa_2$ is given by~(\ref{special_case}).
		The dashed diagonal line in the main figure is $q_{cons}$, discussed under~(\ref{beta}).
		Inset (b) shows how an  unstable solution in the conservative case, marked by a blue cross on the main panel, can be stabilized by the loss and gain (the green domain for $0.32\lesssim\gamma_1\lesssim  0.45$).
		Inset (a) shows examples of a branch of solutions for $0.49\leq\gamma_1\leq 0.99$ situated below $q_{cons}$, at the (black) asterisk in the main panel.
	The (red) green dots in the insets are for (un-) stable solutions (up to $z_m=10^2$).
		The step size in the numerics was $dz=10^{-3}$ and the transverse domain $|x|<50$ ($dx=0.1$). 
		The resolution for the individual calculations in the main panel was $\Delta \kappa_2 = \Delta q = 0.02$. 
		}
	\label{fig:one}
\end{figure}

A large diversity of the particular solutions of the system~(\ref{reduced_equations}) can be found~\cite{MST94}.  
To restrict the number of cases below we concentrate on the simplest ones and start with the $\PT$-symmetric solutions. 
Using the well-known~\cite{KaramzinSukhorukov} soliton of (\ref{reduced_equations}) we obtain that such $\PT$-symmetric soliton exists subject to the constraint (\ref{special_case}), when
$
\gamma_1=\kappa_1\sin\left(\delta \right)$ and $ \gamma_2=\kappa_2\sin(2\delta),
$
and read
\begin{eqnarray}
\label{KarSukh}
\begin{array}{ll}
u_1= 3\beta^2 e^{ipz}/\cosh^2(\beta x) , & u_2=\pm u_1e^{\mp i\delta},
\\
v_1 =3\beta^2 e^{2ipz }/\cosh^2(\beta x), & v_2=\pm v_1e^{\mp 2i\delta},
\end{array}
\end{eqnarray}
where
\begin{eqnarray}
\label{p}
p=\frac{1}{3}\left[2q+\kappa_1\cos(\delta) -2\kappa_2\cos(2\delta)\right],
\\
\label{beta}
\beta^2=\frac{1}{6}\left[q+2\kappa_1\cos(\delta)-\kappa_2\cos(2\delta)\right].
\end{eqnarray}
The right hand side of the last equation must be positive, that imposes the constraint 
on the mismatch of the propagation constants:  
$
q>q_0(\delta)=\kappa_2\cos(2\delta)-2\kappa_1\cos(\delta).
$
For the conservative case $\delta=0$ we have that $q_0(\delta=0)=q_{\rm cons}=\kappa_2-2\kappa_1$. Furthermore one can verify that $q_0(\delta)<q_{\rm cons}$ for the interval $\delta\in(0,\pi/2)$, i.e. the gain and loss introduced in the system reduce the lower band for $q$ for which the exact solution~(\ref{KarSukh}) exists, see inset~(a) of Fig.~\ref{fig:one} for an illustration.
For all other $\delta$ (i.e. for $\delta\in (\pi/2,\pi)$) one has that $q_0>q_{\rm cons}$.

The numerical stability analysis was performed by studying the evolution of initially perturbed stationary states according to~(\ref{basic_equations}) along $z$ and investigating signals of instability. 
The perturbation is invoked at $z=0$ by multiplying the initial conditions for each component of the vector $\psi$ introduced above by the factor $(1+10^{-3}n_j)$, with $j=1,2,3,4$ and $n_j$ being uncorrelated Gaussian random numbers with zero mean and unit variance. 
We then first evaluated to monitor several different signals of instability for threshold values of $10 \%$ deviation from the initial values during the evolution $0 < z \leq z_m =10$ for each of the four modulus of the fields. 
  We numerically found different quantities (and components) to perform qualitatively similar with respect to determining instability. These quantities were: the center of mass; the (maximum) amplitude; and the root mean square (RMS) width. 
Therefore, for transparency, the results on stability presented here (Fig.~\ref{fig:one}) is based on only one signal of instability (for only the first component $u_1$), that is the RMS width.
If no signal of instability occurred, we repeated the analysis with $z_m=10^2$.  
All calculations were done with the $C^{++}$ code generator XMDS~\cite{XMDS1,XMDS2}.
 
 Our main numerical results on the stability of  solutions are summarized in Fig.~\ref{fig:one}. In the main panel, we show the stability of conservative solitons on the diagram ($q,\kappa_2$) for $\kappa_1=1$. 
The solitons exist above the dashed line, which corresponds to the exact analytical solution, i.e. described by (\ref{KarSukh}) with $q=q_{\rm cons}$ and $\delta=0$. 
The solitons in the green stripe
above the dashed line were found stable. All other solitons with larger mismatch of propagation constants are unstable (red domain). 

Turning now to the stability of $\PT$-symmetric solitons given by~(\ref{KarSukh})-(\ref{beta}) 
with gain and loss fulfilling the relation (\ref{special_case}) in the general case ($0 \leq \gamma_1 < \kappa_1 = 1$ and $0 \leq \gamma_2 < \kappa_2$), we show the stability analysis in the two diagrams $(\gamma_1,\gamma_2)$ [insets in Fig.~\ref{fig:one}] for different sets of the "conservative" parameters. 
Gain and loss can stabilize solitons, see inset~(b) of Fig.~\ref{fig:one}, where we show a branch of solutions in the plane ($\gamma_1,\gamma_2$), which bifurcates from an (arbitrarily chosen) unstable conservative soliton marked by a blue cross in the  main panel. 
Hence, we observe that sufficiently large gain and loss can stabilize the solutions (computed stable solitons are shown by green dots). 
Furthermore, since the domain of existence of localized solutions in the presence of gain and loss is larger than that of the conservative case, we considered stability of solitons which do not exist in the conservative limit. 
An illustrative example is shown in inset~(a) of Fig.~\ref{fig:one} and corresponds to the set of parameters indicated by the asterisk on the main panel. 
We again observed that at sufficiently large gain and loss there exists a stability window (green dots).

Such solitons can be observed in a system of tunnel-coupled slab waveguides of LiNbO$_3$~\cite{SBS96,S94}, of a characteristic length $\sim 5$cm. 
For an input beam width $\sim 60\mu$m the diffraction length is $L_d \approx 2.0$mm, corresponding to  $z\approx 25$. 
	A typical linear coupling length is $L_c =\pi/K_j  \approx (1-2) $cm, corresponds to $\kappa_1 \approx \kappa_2 \approx (0.1-0.2)$. 
The absorption and gain induced by active impurity doping can vary in the range $\lesssim 0.17 \ (0.35)$ dBcm$^{-1}$ for the FF (SH), i.e. $\gamma_1 \approx 0.05, \ \gamma_2 \approx 0.1$ in the dimensionless variables. 
Experimentally feasible input powers for solitons generation for quadratic nonlinearity parameter $\chi^{(2)}$= 5.6pm/V is  $\sim  10kW$, corresponding to~$u \sim 1$~(dimensionless).

\begin{figure}[ht!]
	\centering
	\includegraphics[width=40mm]{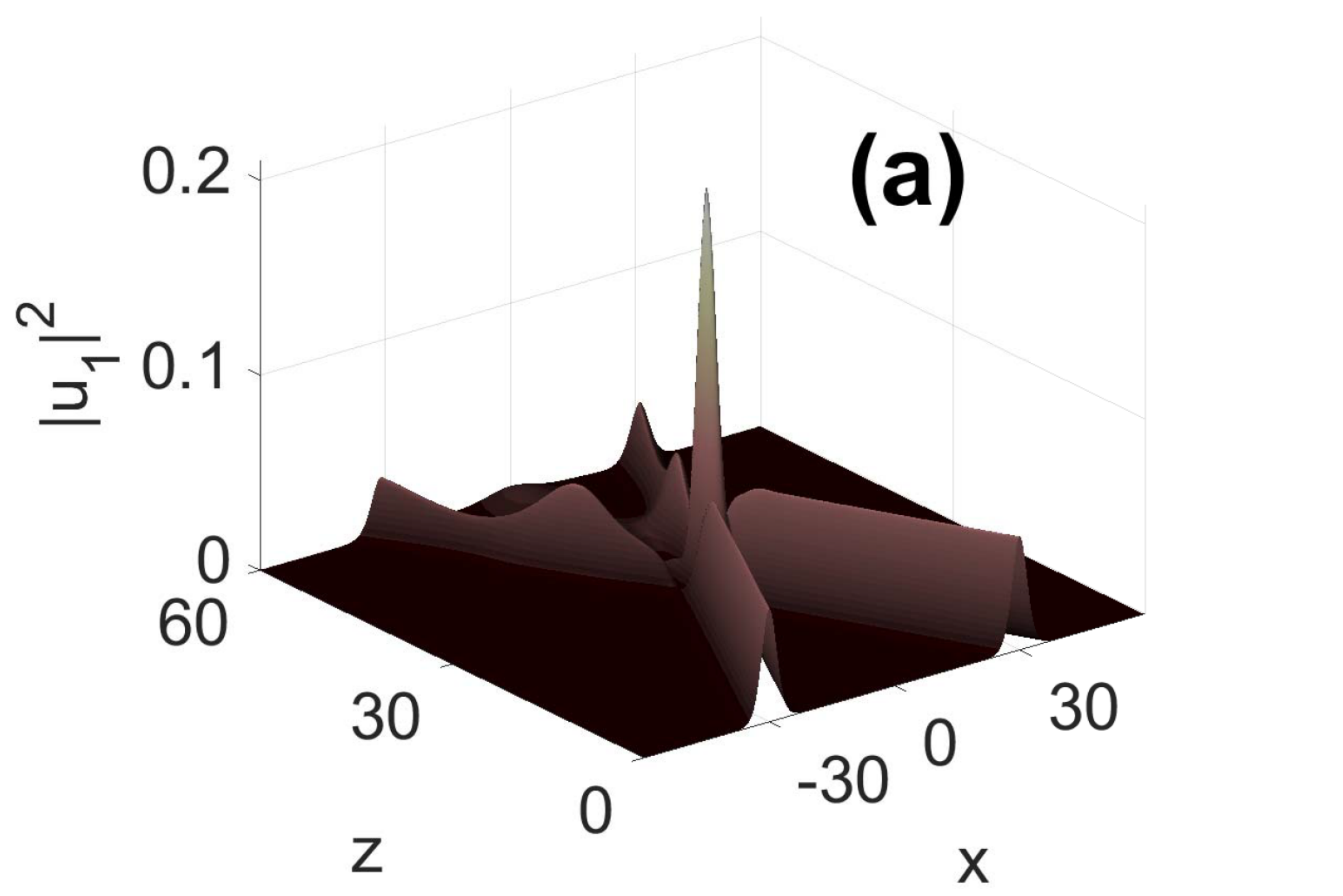}
	\includegraphics[width=40mm]{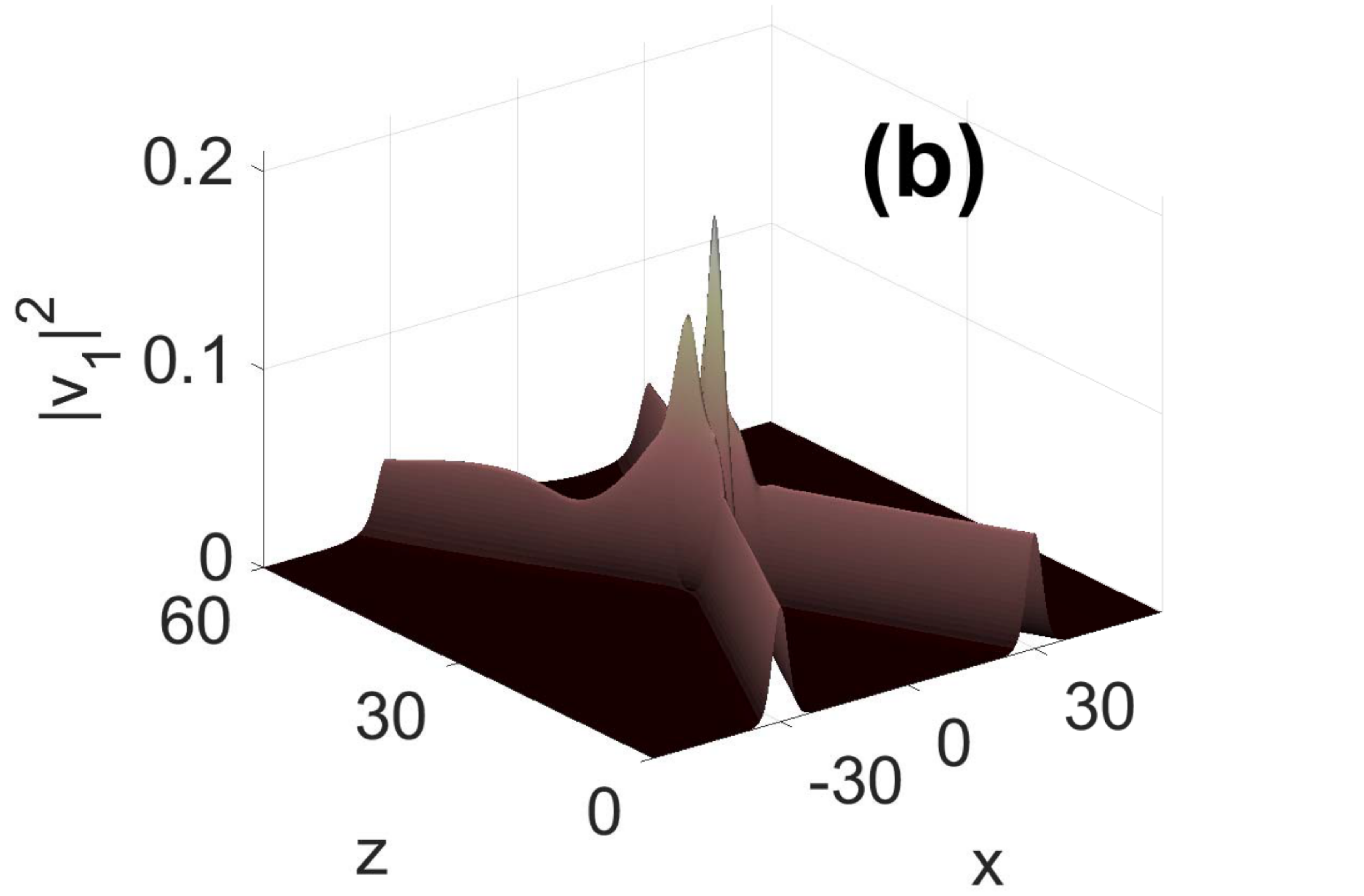}
	\caption{Interaction of two solitons with equal intensity profiles of the FF and SH, given by~(\ref{KarSukh}). Panel (a)/(b) shows the modulus square of the $u_1$-/$v_1$-component. 
	From~(\ref{KarSukh}) we see that $|u_2|^2$ and $|v_2|^2$ are proportional to those.  
	Parameters here were $\kappa_1=1$, $\kappa_2=0.5$, $q=-1$, $\gamma_1=0.5$, $\gamma_2 \simeq 0.433$, i.e. from~(\ref{special_case}), with velocity $w=\pm 1$ for the two initial pulses.}
	\label{fig:two}
\end{figure}

Stable $\PT$-symmetric solitons 
were tested with respect to the mutual interactions, an example is shown in Fig.~\ref{fig:two},
and we verified that the energy is constant ($P_1=P_2=\textnormal{constant}$) when~(\ref{special_case}) is fulfilled. The collision however cannot be seen as strictly elastic, because weak modulation of the pulse shapes after collision is detectable. 

\begin{figure}[ht!]
	\centering
	\includegraphics[width=40mm]{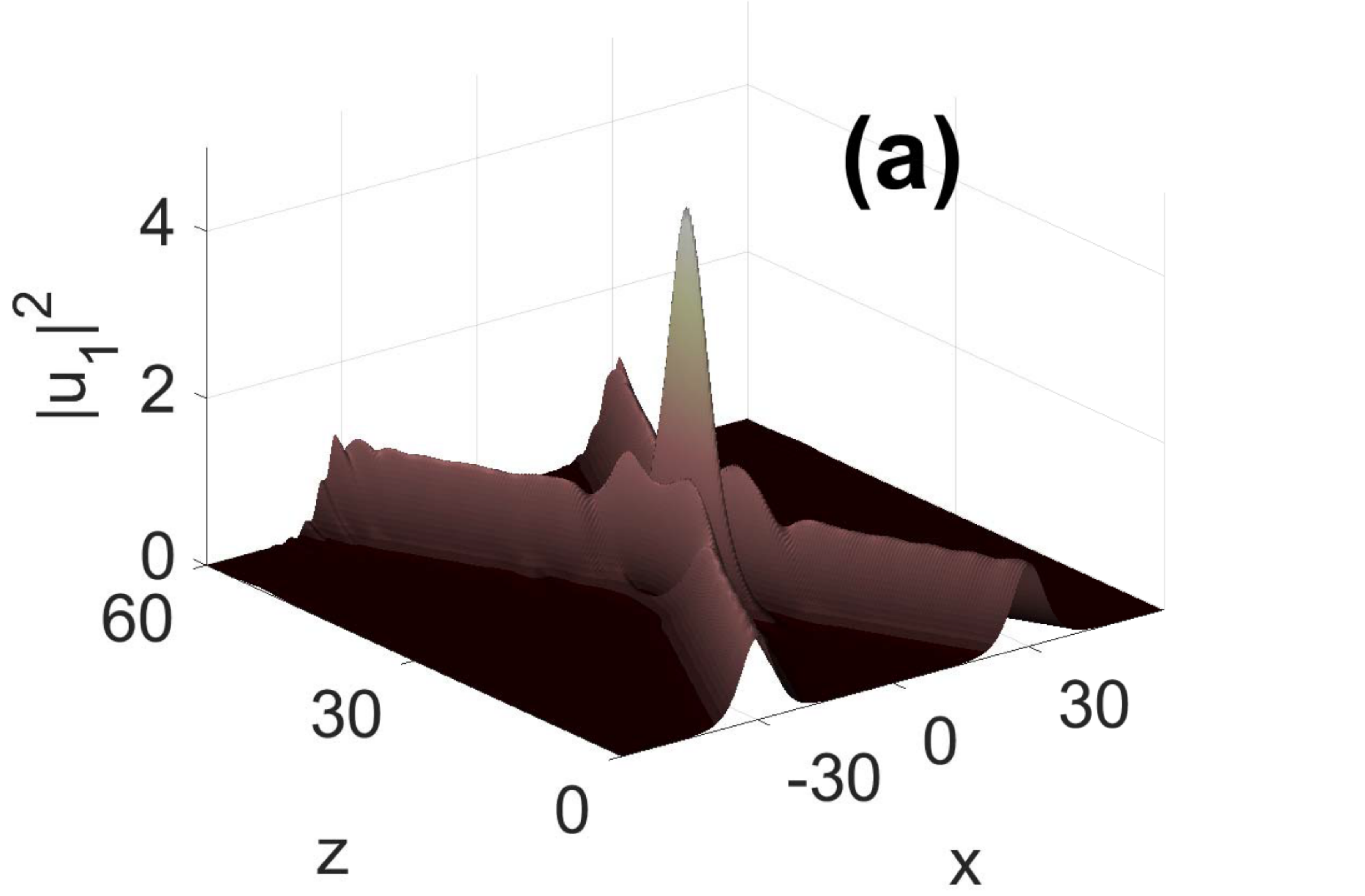}
	\includegraphics[width=40mm]{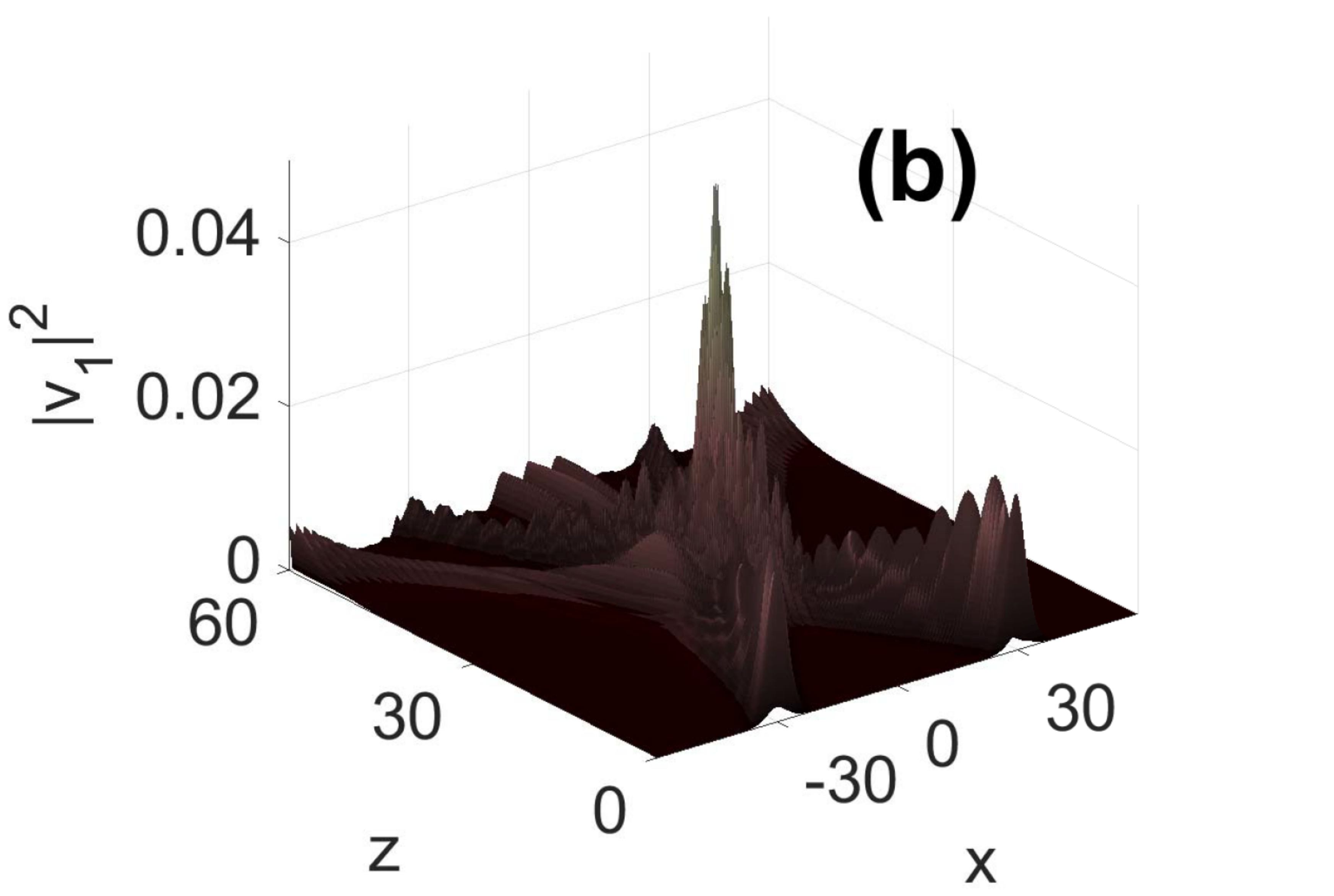}
	\caption{Interaction of two solitons with initial profiles obtained from the cascading limit~(\ref{cascading_approx_1}) and~(\ref{cascad_soliton}).
	The parameters for the initial conditions here were $A=1$, $\kappa_1=1$, $\kappa_2= 20$, $q=-3$, $\gamma_1=\gamma_2=0.1$ (i.e. $\Delta \simeq -400$) and $w=\pm 1$. 
	}
	\label{fig:four}
\end{figure}

In the case of the second-harmonic generation in a single waveguide an approximate solution can be obtained in the so-called cascading limit, which corresponds to the large mismatch parameter $|q| \gg 1$~\cite{Stegeman}. 
This allows one to reduce the description of the two component system to the single nonlinear Schr\"odinger (NLS) equation for the FF only. Similar reduction is also possible in the case of the $\PT$-symmetric coupler~(\ref{basic_equations}). To this end we  introduce $\Delta=q^2+\gamma_2^2-\kappa_2^2$ and require $|\Delta|$ to be large enough. Notice, that this last condition can be satisfied not only due to large mismatch $q$ (as in the conservative systems) but also due to the strong coupling $\kappa_2$ of the SHs. Now the derivatives of $v_{1,2}$ can be neglected and one computes
\begin{equation}
v_1 \approx -\frac{(q+i\gamma_2) u_1^2+\kappa_2u_2^2}{\Delta}, \,\,\,
v_2 \approx -\frac{(q- i\gamma_2)u_2^2 +\kappa_2u_1^2}{\Delta}.
\label{cascading_approx_1}
\end{equation}
The equations for the FF are now reduced to
\begin{equation}
\label{NLS}
\begin{array}{l}
i{u}_{1,z} = - u_{1,xx}+ \kappa_1 u_2+ i\gamma_1 u_1+\frac{2(q+i\gamma_2)}{\Delta}|u_1|^2u_1
+\frac{2\kappa_2}{\Delta} u_1^*u_2^2,
\\
i{u}_{2,z} = - u_{2,xx}+ \kappa_1 u_1- i\gamma_1 u_2 +\frac{2(q-i\gamma_2)}{\Delta}|u_2|^2u_2
+\frac{2\kappa_2}{\Delta}u_2^*u_1^2.
\end{array}
\end{equation}
Thus we obtained a $\PT$-symmetric coupler with self-phase modulation and the four wave mixing terms due to coupling of the SHs, as well as with linear and nonlinear gain and losses. 
At $\gamma_2=0$, (\ref{NLS}) is reduced to the $\PT$-symmetric dimer model~\cite{AKOS,DrbMal}, while for an $x$-independent plane wave solution~(\ref{NLS}) becomes the nonlinear coupler~\cite{MMK}.
Both limits have been studied in the literature (see e.g.~\cite{review} and references therein). The gain and loss in the second component, thus, introduce nonlinear gain and loss for the first component. 
An interesting feature of the cascading limit (\ref{NLS}) is that the sign of its effective Kerr-like nonlinearity is determined by the coefficient $q/\Delta$ and hence can be either focusing ($q/\Delta<0$) or defocusing ($q/\Delta>0$). 
A soliton solution of (\ref{NLS}) is readily found in the form
\begin{eqnarray}
\label{cascad_soliton}
u_{1} = {A}{ \mbox{sech}(Ax/\sqrt{\kappa_2 \cos(2\delta)-q} ) }, \quad u_2=\pm u_1e^{\mp i\delta}, 
\end{eqnarray}
provided the condition~(\ref{special_case}) holds and the additional constraint $q<\kappa_2\cos(2\delta)$ is satisfied.
After having numerically confirmed the stability of the approximate solutions~(\ref{cascading_approx_1}) and~(\ref{cascad_soliton}) (given that $|\Delta| \gg 1$ and small $\gamma_{1,2}$),
we have tested also those solutions with respect to mutual interactions, see Fig.~\ref{fig:four} for an example.
We have verified that we have solitons for a large domain of initial conditions fulfilling $|\Delta| \gg 1$, though generally with oscillating energies, $P_1$ and $P_2$ in the two waveguides (with $P_1 + P_2 \neq \textnormal{constant}$), unless the condition~(\ref{special_case}) is fulfilled. 

\begin{figure}[ht!]
	\centering
	\includegraphics[width=40mm]{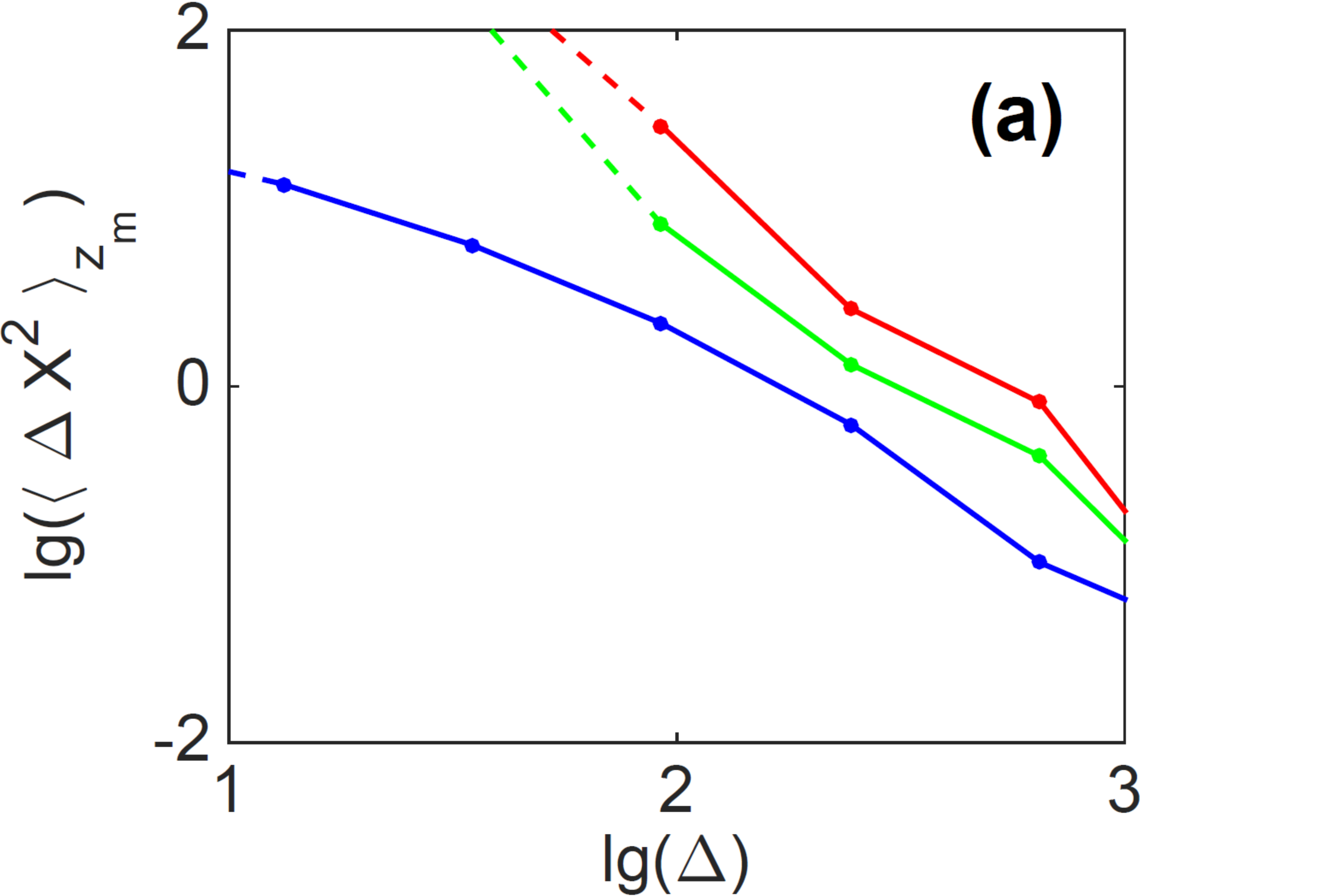}
	\includegraphics[width=40mm]{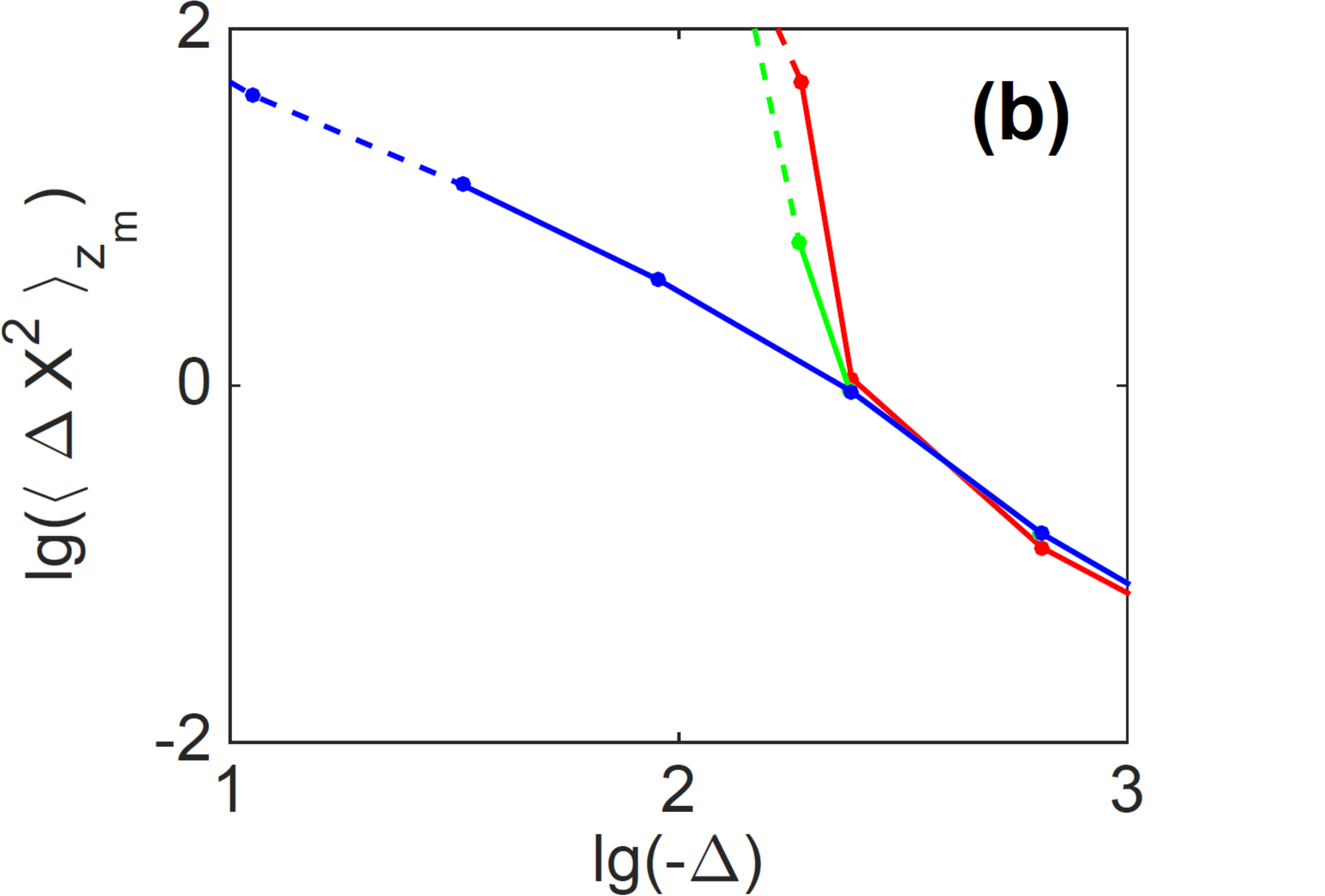}
	\caption{Characterization of the solutions in the cascading limit, using (\ref{cascading_approx_1}) and~(\ref{cascad_soliton}) for the initial conditions. 
In (a) ($\Delta>0$) $q$ was varied while $\kappa_2=1$ ($\kappa_1=1$).
In (b) ($\Delta<0$) $\kappa_2$ was varied while $q=1$.
Blue curves
are for $\gamma_1=\gamma_2=0$, red curves are for $\gamma_1=\gamma_2=0.05$. The green curve in (a) is for $\gamma_1=0.05$ and $\gamma_2\simeq 0.0999$ according to Eq.~(\ref{special_case}).
When varying $\kappa_2$ ($\Delta<0$) the use of~(\ref{special_case}) leads to different values of $\gamma_2$, the resulting green curve in (b) is partly overlapping the red. 
For the data points with dashed curves to the left, a drift along $x$ was notable. 
}
\label{figure4}
\end{figure}

To quantify how~(\ref{cascading_approx_1}) and~(\ref{cascad_soliton}) works for different values of $\Delta$, we define the following numerical measure
\begin{equation}
\langle \Delta X^2 \rangle_{z_{m}} \equiv    {  \langle |X^{2}(z) - X^{2}(0)| \rangle_{z_m}}\bigg/ {X^{2}(0)},   
\label{def_measure}
\end{equation}
where 
$\langle f\rangle_{z_m}=z_{m}^{-1} \int_0^{z_{m}} f(z)dz$ with $X^{2}(z) = P_1^{-1} \int  x^2 \left( |u_1|^2 + 2|v_1|^2 \right) dx.$
Here $z_{m}$ should be chosen large enough, such that $\langle \Delta X^2 \rangle_{z_{m}}$ is qualitatively independent of~$z_{m}$. 
The characterization of the solutions using the parameter $\langle \Delta X^2 \rangle_{z_{m}}$ is shown in Fig.~\ref{figure4}, where we used $z_{m}=10^4$ ($dz =10^{-3}$) and $|x|<10^2$ ($dx =0.2$) for all curves. 
We observe that for the parameter $\Delta$, in spite of the fact that it is combined of three system parameters (due to (\ref{special_case}) only two of them are independent), the curves for different $q$ and $\kappa_2$ indicate the same qualitative behavior in the cascading limit, 
and that the average deformation of the soliton shape~(\ref{def_measure}) decreases fast with $|\Delta|$. 
This numerically confirms the validity of the approach.

To conclude, we have obtained two families of stable $\PT$- symmetric solitons in a $\PT$- symmetric $\chi^{(2)}$ coupler.
We numerically found, in the case of solitons with equal shapes of the first and second harmonics, a region of stability for the solitons.
It is established that gain and loss can increase the domain of soliton existence with respect to the propagation constant mismatch and can stabilize solitons which are unstable in the conservative limit. 
Stable solitons interact nearly elastically. 
We also analyzed the cascading limit, which is reduced to a $\PT$-symmetric dimer with linear and nonlinear gain and loss. 
Solitons in this limit can still be found stable,
but their interactions manifest appreciable non-elastic effects.

VVK acknowledges 
hospitality of \"{O}rebro University (Sweden) and International Islamic University (Malaysia). F.A. was supported by the grant FRGS 16-014-0513 (IIUM).


\begin{thebibliography}{99}   

\bibitem{review1}
C. Etrich, F. Lederer, B. A. Malomed, T. Peschel, and U. Peschel,
Prog. Opt. {\bf 41}, 483-568 (2000).

\bibitem{review2}
A. V. Buryak, P. Di Trapani, D. V. Skryabin, and S. Trillo,
Phys. Rep. {\bf 370}, 63-235 (2002).

\bibitem{KaramzinSukhorukov} Y. N. Karamzin, and A. P. Sukhorukov,
Sov. Phys.-JETP. {\bf 41}, 414 (1976).

\bibitem{WD94a}
M. J. Werner and P. D. Drummond,
Opt. Lett. {\bf 19}, 613-615 (1994).

\bibitem{WD94b}
M. J. Werner and P. D. Drummond,
J. Opt. Soc. Am. B {\bf 10}, 2390 (1993).

\bibitem{MST94}
C. R. Menyuk, R. Schiek, and L. Torner,
J. Opt. Soc. Am. B {\bf11}, 2434 (1994).

\bibitem{BK95}
A.V. Buryak and  Yu. S. Kivshar,
Opt. Lett. {\bf 19}, 1612-1614 (1994).

\bibitem{SBS96}
R. Schiek, Y. Baek, and G. Stegeman,
Phys. Rev. E {\bf 53} , 1138 (1996).

\bibitem{SBSS99}
R. Schiek, Y. Baek,  G. Stegeman, and W. Sohler,
Opt. Lett.  {\bf 24} , 83-85 (1999).

\bibitem{Iwanow}
	R. Iwanow, R. Schiek, G. Stegeman, T. Pertsch, F. Lederer, Y. Min, and W. Sohler,
Opto-electronics review, {\bf 13}, 113 (2005).

\bibitem{Stegeman}
G. I. Stegeman, D. J. Hagan, L.  Torner,
Opt. and Quant. Electron. {\bf 28}, 1691-1740 (1996).

\bibitem{Polyak}
Mirosław A. Karpierz,
Opt. Appl., {\bf 26}, 391-396 (1996).

\bibitem{mak97}
W. C. K. Mak, B. A. Malomed, and P. L. Chu,
Phys. Rev. E {\bf 55}, 6134 (1997).
 
\bibitem{Bender}
C. M. Bender,
Rep.  Progr.  Phys.  {\bf 70}, 947–1018 (2007)

\bibitem{review}  V. V. Konotop, J. Yang, and D. A. Zezyulin,
Rev. Mod. Phys.  {\bf 88}, 035002 (2016).

\bibitem{Sukho} D. A. Antonosyan, A. S. Solntsev,  and A. A. Sukhorukov,    
Opt. Lett. {\bf 40}, 4575-4578 (2015). 

\bibitem{El} R. El-Ganainy,  J. I. Dadap,  and R. M. Osgood,   
Opt. Lett. {\bf 40}, 5086–5089 (2015). 

\bibitem{Tripen}
T. Wasak,  P. Sza\'nkowski, V. V. Konotop, and M. Trippenbach,  
Opt. Lett. {\bf 40}, 5291-5294, (2015).  
 
\bibitem{chi2sol}
K. Li, D. A. Zezyulin, P. G. Kevrekidis, V. V. Konotop, and F. Kh. Abdullaev,
Phys. Rev. A {\bf 88}, 053820 (2013).

\bibitem{DrbMal} Driben, R., and B. A. Malomed,
Opt. Lett., {\bf 36}, 4323–4325 (2011).

\bibitem{XMDS1} G. R. Collecutt and P. D. Drummond,
 Comput. Phys. Commun. {\bf 142}, 219 (2001).

\bibitem{XMDS2} G. R. Dennis, J. J. Hope and M. T. Johnsson,
 Comput. Phys. Commun. {\bf 184}, 201 (2013).

\bibitem{AKOS} F. Kh. Abdullaev, V. V. Konotop, M. \"{O}gren, M.P. S{\o}rensen,
 Opt. Lett. {\bf 36}, 4566-4568 (2011).

\bibitem{MMK} A. E. Miroshnichenko, B. A. Malomed, and Y. S. Kivshar,
Phys. Rev. A {\bf 84}, 012123 (2011).
 
\bibitem{S94}
R. Schiek, Y. Baek, G. Krijnen, G. I. Stegeman, I. Baumann and W. Sohler,
Opt. Lett. {\bf 21}, 940-942 (1996).
 
\end{thebibliography}
\end{document}